# On the uniform tiling with electrical resistors

*M. Q. Owaidat*

*Department of Physics, Al-Hussein Bin Talal University, Ma'an,71111, Jordan*

*E-mail: Owaidat@ahu.edu.jo*

**Abstract**

We calculate the effective resistance between two arbitrary lattice points on infinite strip of the triangular lattice (ladder network) in one dimension, and on infinite modified square and Union Jack lattices in two dimensions, and on infinite decorated simple cubic and base-centered cubic lattices in three dimensions by using the general lattice Green's function method.

**Keywords:** lattice Green's function; modified square ;Union Jack lattice; base-centered lattice.

 **PACS numbers**: 01.55.+b, 02.30.Vv, 41.20.Cv

## 1. Introduction

The determination of the effective resistance between any two nodes on an infinite lattice network of identical resistors is a problem of considerable interest in electric circuit theory. The problem has been studied extensively by many authors using several methods[1-6]. Recently, Cserti [6] has presented an elegant method to calculate the resistance between two arbitrary nodes for infinite lattice networks. This approach is based on the lattice Green's function of the Laplacian operator of the difference equations governed by the Ohm's and Kirchhoff's laws.

Cserti et al. [7] have applied the Green's function method to the problem of a perturbed network that is obtained by removing one bond from the perfect lattice. Based on this method, other perturbations are considered such as replacing one resistor in the perfect lattice with another one[8], introducing single interstitial resistor[9] and two extra interstitial resistors [10] between two arbitrary nodes in the perfect lattice. The Green's function method is also a useful tool to calculate the capacitance of a perfect and a perturbed infinite capacitor networks[11-13].

Another interesting problem in electric circuit theory is to compute the two-point resistance in a finite lattice network. Recently, Wu[14] has presented a general formulation for computing two-point resistances in a network in terms of the eigenvalues and eigenfunctions of the Laplacian matrix. Jafarizadeh et al.[15] proposed an algorithm for the calculation of the effective resistance between two arbitrary nodes in an arbitrary distance-regular networks.

Cserti et al. [16] have generalized the Green's function method [6] for a resistor network that is a uniform tiling of d-dimensional space with electrical resistors. They presented several examples as applications to this method. Owaidat[17] has used the general Green's function method [16] to calculate the two-point resistance on the face-centered cubic lattice.

In this paper the same technique [16] is applied to other lattices: infinite full ladder, modified square, Union Jack, decorated simple cubic and base-centered cubic lattices of identical resistors. The paper is arranged as follows. In section 2, a review of general formalism for computing the two point resistance on an any infinite lattice network is given[17]. In section 3, several lattice structures in one, two and three dimensions with more one type of lattice point with some analytical and numerical results for resistance are presented.. A conclusion is given in section 4. The results for the resistance on a decorated simple cubic lattice in terms of the resistances on an usual simple cubic lattice are listed in Appendix.

**2. Review the general formulation**

The formulation of calculating the resistance between two arbitrary sites in an infinite uniform lattice network of resistors is briefly reviewed (see, for more details Ref.16). Consider a regular d- dimensional lattice comprised of repeated unit cells, each contains $v$ vertices(sites) labeled by $\alpha = 1, 2, ..., v$ ,specified by lattice site $\mathbf{r}_n = \sum_{i=1}^{d} n_i \mathbf{a}_i$, where $\mathbf{a}_1, \mathbf{a}_2, ..., \mathbf{a}_d$ are the unit cell vectors and $n_1, n_2, ..., n_d$ are integers. If $\{\mathbf{r}_n, \alpha\}$ denotes any lattice site, then $I_\alpha(\mathbf{r}_n)$ and $V_\alpha(\mathbf{r}_n)$ denote the current and potential at site $\{\mathbf{r}_n, \alpha\}$, respectively. It is supposed that the current can enter the $\{\mathbf{r}_n, \alpha\}$ site from a source outside the lattice. According to Ohm's and Kirchhoff's laws the equations for the currents $I_\alpha(\mathbf{r}_n)$ for each lattice site in one unit cell can be written as

$$\sum_{m,\beta} L_{\alpha\beta}(\mathbf{r}_m - \mathbf{r}_n) V(\mathbf{r}_m) = -I_\alpha(\mathbf{r}_n) \qquad (2.1)$$

where $L_{\alpha\beta}(\mathbf{r}_m - \mathbf{r}_n)$ is a $v$ by $v$ matrix called Laplacian matrix. The Laplacian matrix .In matrix notation one writes Eq. (2.1) as

$$\sum_{m} \mathbf{L}(\mathbf{r}_m - \mathbf{r}_n) \mathbf{V}(\mathbf{r}_m) = -\mathbf{I}(\mathbf{r}_n) \qquad (2.2)$$

The discrete Fourier transforms of the current and potential are defined as

$$I_\alpha(\mathbf{k}) = \sum_{\mathbf{r}} I_\alpha(\mathbf{r}) e^{-i\mathbf{k}\mathbf{r}} \quad , \quad V_\alpha(\mathbf{k}) = \sum_{\ell} V_\alpha(\mathbf{r}_\ell) e^{-i\mathbf{k}\mathbf{r}_\ell} \quad , \quad \alpha = 1, 2, ..., v \qquad (2.3)$$

where **k** is the wave vector in the reciprocal lattice and is limited to the first Brillouin zone[18-21]. The general expressions for the inverse Fourier transform are given

$$I_\alpha(\mathbf{r}_\ell) = \mathfrak{I}^{-1}[I_\alpha(\mathbf{k})] = \frac{\Omega_0}{(2\pi)^d} \int_{-\pi/a_1}^{\pi/a_1} ,..., \int_{-\pi/a_d}^{\pi/a_d} I_\alpha(\mathbf{k}) e^{i\mathbf{k}\mathbf{r}_\ell} d\mathbf{k} \qquad (2.5)$$

$$V_\alpha(\mathbf{r}_\ell) = \mathfrak{I}^{-1}[V_\alpha(\mathbf{k})] = \frac{\Omega_0}{(2\pi)^d} \int_{-\pi/a_1}^{\pi/a_1} ,..., \int_{-\pi/a_d}^{\pi/a_d} V_\alpha(\mathbf{k}) e^{i\mathbf{k}\mathbf{r}_\ell} d\mathbf{k} \qquad (2.6)$$

where d is the dimensionality and $\Omega_0$ is the volume of the unit cell. Using the above equations, Eq. (2.2) can be written as

$$\mathbf{L}(\mathbf{k})\mathbf{V}(\mathbf{k}) = -\mathbf{I}(\mathbf{k}) \qquad (2.7)$$

where $\mathbf{L}(\mathbf{k})$ is the Fourier transform of Laplacian matrix and given by

$$\mathbf{L}(\mathbf{k}) = \sum_\ell \mathbf{L}(\mathbf{r}_\ell) e^{-i\mathbf{k}\mathbf{r}_\ell} \qquad (2.8)$$

The lattice Green's function $\mathbf{G}(\mathbf{k})$ (which is $v$ by $v$ matrix) corresponding to Fourier transform of Laplacian matrix $\mathbf{L}(\mathbf{k})$ is defined as

$$\mathbf{G}(\mathbf{k}) = -\mathbf{L}^{-1}(\mathbf{k}) \qquad (2.9)$$

Hence, Eq. (2.7) becomes

$$\mathbf{V}(\mathbf{k}) = \mathbf{G}(\mathbf{k})Q(\mathbf{k}) \qquad (2.10)$$

To compute the resistance $R_{\alpha\beta}(\mathbf{r}_m, \mathbf{r}_n)$ between two nodes $\{\mathbf{r}_m, \alpha\}$ and $\{\mathbf{r}_n, \beta\}$, we connect $\{\mathbf{r}_m, \alpha\}$ and $\{\mathbf{r}_n, \beta\}$ to the two terminals of an external battery and measure the current $I$ going through the battery while no other nodes are connected to external sources. Let the potentials at the two nodes be, respectively, $V_\alpha(\mathbf{r}_m)$ and $V_\beta(\mathbf{r}_n)$. Then, the desired resistance is the ratio

$$R_{\alpha\beta}(\mathbf{r}_m,\mathbf{r}_n) = \frac{V_\alpha(\mathbf{r}_m) - V_\beta(\mathbf{r}_n)}{I} \tag{2.11}$$

The computation of the two-point resistance $R_{\alpha\beta}(\mathbf{r}_m,\mathbf{r}_n)$ is now reduced to solving (2.1) or (2.6) for $V_\alpha(\mathbf{r}_m)$ and $V_\beta(\mathbf{r}_n)$ with the current given by

$$I_\lambda(\mathbf{r}_k) = I\left(\delta_{\mathbf{r}_k,\mathbf{r}_m}\delta_{\lambda,\alpha} - \delta_{\mathbf{r}_k,\mathbf{r}_n}\delta_{\lambda,\beta}\right) \tag{2.12}$$

Therefore, combining Eqs.(2.6),(2.10) and (2.12) one can obtain the following integral expression for the electrical potential in any site of the lattice network:

$$V_\mu(\mathbf{r}_k) = \frac{\Omega_0}{(2\pi)^d} \int_{-\pi/a_1}^{\pi/a_1} ,...,\int_{-\pi/a_d}^{\pi/a_d} \left(G_{\mu\alpha}(\mathbf{k})e^{-i\mathbf{k}\mathbf{r}_m} - G_{\mu\beta}(\mathbf{k})e^{-i\mathbf{k}\mathbf{r}_n}\right)e^{i\mathbf{k}\mathbf{r}_k} d\mathbf{k} \tag{2.13}$$

Substituting potential distribution given Eq.(2.13) into Eq. (2.11), the node-node resistance can be written as follows:

$$R_{\alpha\beta}(\mathbf{r}_m,\mathbf{r}_n) = \frac{\Omega_0}{(2\pi)^d} \int_{-\pi/a_1}^{\pi/a_1},...,\int_{-\pi/a_d}^{\pi/a_d} \left(G_{\alpha\alpha}(\mathbf{k}) + G_{\beta\beta}(\mathbf{k}) - G_{\alpha\beta}(\mathbf{k})e^{-i\mathbf{k}(\mathbf{r}_n-\mathbf{r}_m)} - G_{\beta\alpha}(\mathbf{k})e^{i\mathbf{k}(\mathbf{r}_n-\mathbf{r}_m)}\right) d\mathbf{k} \tag{2.14}$$

The resistance does not change if the lattice structure is deformed and the topology of the lattice structure is preserved. Therefore, any resistor network lattice can be deformed into a d-dimensional hypercube where the volume of the unit cell is $\Omega_0 = a_1 a_2 ... a_d$. If one writes $\mathbf{r}_n - \mathbf{r}_m = \mathbf{r}_\ell = \ell_1 \mathbf{a}_1 + ... + \mathbf{a}_d \ell_d$ and changes the variables, $\mathbf{k} a_i = \theta_i (i=1,2,...,d)$ in equation (2.14) then the final result for the resistance between any infinite uniform lattice structure can be written as

$$R_{\alpha\beta}(\ell_1,...,\ell_d) = \int_{-\pi}^{\pi}\frac{d\theta_1}{2\pi}..\int_{-\pi}^{\pi}\frac{d\theta_d}{2\pi}\left\{G_{\alpha\alpha}(\theta_1,..\theta_d) + G_{\beta\beta}(\theta_1,..\theta_d) \right.$$
$$\left. -G_{\alpha\beta}(\theta_1,..\theta_d)e^{-i(\ell_1\theta_1+..+\theta_d\ell_d)} - G_{\beta\alpha}(\theta_1,..\theta_d)e^{i(\ell_1\theta_1+..+\theta_d\ell_d)}\right\} \tag{2.15}$$

## 3. Applications

### 3.1. One- dimensional lattice

Here we consider a strip of the triangular lattice of resistors as an example of a one - dimensional resistor network with next-nearest- neighbor resistors as shown in Fig.1. We assume all edges represent a resistance R. Here one can choose a unit cell contains two lattice points labeled by 1 and 2. According to Ohm's and Kirchhoff's laws the currents in the unit cell at a lattice point $x = \ell a$ are given by

$$\left.\begin{array}{l} I_1(x)R = 4V_1(x) - V_1(x+a) - V_1(x-a) - V_2(x) - V_2(x+a) \\ I_2(x)R = 4V_2(x) - V_2(x+a) - V_2(x-a) - V_1(x) - V_1(x-a) \end{array}\right\} \quad (3.1)$$

The Laplacian matrix is two by two and can be written as

$$\mathbf{L}(x) = \frac{1}{R}\begin{bmatrix} -4\delta_{x,0} + \delta_{x,a} + \delta_{x,-a} & \delta_{x,0} + \delta_{x,-a} \\ \delta_{x,0} + \delta_{x,a} & -4\delta_{x,0} + \delta_{x,a} + \delta_{x,-a} \end{bmatrix} \quad (3.2)$$

and after changing the variable $ka = \theta$, the Fourier transformation reads

$$\mathbf{L}(\theta) = \frac{1}{R}\begin{bmatrix} 2\cos\theta - 4 & 1+e^{i\theta} \\ 1+e^{-i\theta} & 2\cos\theta - 4 \end{bmatrix} \quad (3.3)$$

The lattice Green's function can be evaluated readily using Eq.(2.9). The result which we want is

$$\mathbf{G}(\theta) = \frac{1}{\text{Det}(\mathbf{L}(\theta))}\begin{bmatrix} 4 - 2\cos\theta & 1+e^{i\theta} \\ 1+e^{-i\theta} & 4 - 2\cos\theta \end{bmatrix} \quad (3.4)$$

where $\text{Det}(\mathbf{L}(\theta)) = (14 - 18\cos\theta + 4\cos^2\theta)/R$ is the determinant of matrix $\mathbf{L}(\theta)$. The resistance between the origin and site $(\ell)$ can be calculated from equation (2.15):

$$R_{\alpha\beta}(\ell) = \int_{-\pi}^{\pi}\frac{d\theta}{2\pi}\left\{G_{\alpha\alpha}(\theta) + G_{\beta\beta}(\theta) - G_{\alpha\beta}(\theta)e^{-i\ell\theta} - G_{\beta\alpha}(\theta)e^{i\ell\theta}\right\} \quad (3.5)$$

There are four types of resistance:

$$R_{11}(\ell) = R_{22}(\ell) = 2R\int_{-\pi}^{\pi} \frac{d\theta}{2\pi} \frac{(2-\cos\theta)(1-\cos\ell\theta)}{(7-2\cos\theta)(1-\cos\theta)} \tag{3.6a}$$

$$R_{12}(\ell) = R\int_{-\pi}^{\pi} \frac{d\theta}{2\pi} \frac{4-2\cos\theta-\cos\ell\theta-\cos(\ell-1)\theta}{(7-2\cos\theta)(1-\cos\theta)} \tag{3.6b}$$

$$R_{21}(\ell) = R\int_{-\pi}^{\pi} \frac{d\theta}{2\pi} \frac{4-2\cos\theta-\cos\ell\theta-\cos(\ell+1)\theta}{(7-2\cos\theta)(1-\cos\theta)} \tag{3.6c}$$

Using above equations, some analytical results(in units of R) are listed below:

$$R_{12}(0) = R_{21}(0) = R_{12}(1) = R_{21}(-1) = \frac{1}{\sqrt{5}},$$

$$R_{11}(1) = R_{22}(1) = \frac{\sqrt{5}-1}{\sqrt{5}}, \quad R_{12}(2) = \frac{4-\sqrt{5}}{\sqrt{5}},$$

$$R_{11}(2) = R_{22}(2) = \frac{5\sqrt{5}-9}{\sqrt{5}}$$

**3.2. Two –dimensional lattices**

In this subsection, we consider two –dimensional lattices with more than one type of site (the square, triangular and honeycomb lattices have been studied in Refs.[5,6] ; the centered square, Kagomé, dice, decorated square and 4.8.8(bathroom tile) lattices in Ref.[16]).

*3.2. 1.Modified square lattice*

The modified square lattice of the resistor network is shown in Fig.2, where a unit cell contains $s = 4$ sites numbered by $\alpha = 1, 2, 3, 4$, and the coordination number is 6. Using Ohm's and Kirchhoff's laws the currents at the lattice site $\{\mathbf{r}, \alpha\}$ with ( $\alpha = 1, 2, 3, 4$) are given by

$$\left.\begin{aligned}RI_1(\mathbf{r}) &= 6V_1(\mathbf{r}) - V_2(\mathbf{r}) - V_2(\mathbf{r}-\mathbf{a}_1) - V_3(\mathbf{r}-\mathbf{a}_1) - V_3(\mathbf{r}-\mathbf{a}_2) - V_4(\mathbf{r}) - V_4(\mathbf{r}-\mathbf{a}_2) \\ RI_2(\mathbf{r}) &= 6V_2(\mathbf{r}) - V_1(\mathbf{r}) - V_1(\mathbf{r}+\mathbf{a}_1) - V_3(\mathbf{r}) - V_3(\mathbf{r}-\mathbf{a}_2) - V_4(\mathbf{r}+\mathbf{a}_1) - V_4(\mathbf{r}-\mathbf{a}_2) \\ RI_3(\mathbf{r}) &= 6V_3(\mathbf{r}) - V_1(\mathbf{r}+\mathbf{a}_1) - V_1(\mathbf{r}+\mathbf{a}_2) - V_2(\mathbf{r}) - V_2(\mathbf{r}+\mathbf{a}_2) - V_4(\mathbf{r}) - V_4(\mathbf{r}+\mathbf{a}_1) \\ RI_4(\mathbf{r}) &= 6V_4(\mathbf{r}) - V_1(\mathbf{r}) - V_1(\mathbf{r}+\mathbf{a}_2) - V_2(\mathbf{r}-\mathbf{a}_1) - V_2(\mathbf{r}+\mathbf{a}_2) - V_3(\mathbf{r}) - V_3(\mathbf{r}-\mathbf{a}_1)\end{aligned}\right\} \quad (3.7)$$

Hence, after changing the variables the Fourier transform of the Laplacian matrix is given by

$$R\mathbf{L}(\theta_1,\theta_2) = \begin{bmatrix} -6 & 1+e^{i\theta_1} & e^{i\theta_1}+e^{i\theta_2} & 1+e^{i\theta_2} \\ 1+e^{-i\theta_1} & -6 & 1+e^{i\theta_2} & e^{-i\theta_1}+e^{i\theta_2} \\ e^{-i\theta_1}+e^{-i\theta_2} & 1+e^{-i\theta_2} & -6 & 1+e^{-i\theta_1} \\ 1+e^{-i\theta_2} & e^{i\theta_1}+e^{-i\theta_2} & 1+e^{i\theta_1} & -6 \end{bmatrix} \quad (3.8)$$

and the Green function can be obtained from Eq. (2.9):

$$\mathbf{G}(\theta_1,\theta_2) = \frac{1}{Det(\mathbf{L}(\theta_1,\theta_2))}\begin{bmatrix} A & C & D & E \\ C^* & B & E & F \\ D^* & E^* & A & C^* \\ E^* & F^* & C & B \end{bmatrix} \quad (3.9)$$

where $Det(\mathbf{L}(\theta_1,\theta_2)) = 256(3-\cos\theta_1-\cos\theta_2-\cos\theta_1\cos\theta_2)/R$ is the determinant of matrix $\mathbf{L}(\theta_1,\theta_2)$, $A = 176-16\cos\theta_1-16\cos\theta_2-16\cos(\theta_1+\theta_2)$,

$B = 176-16\cos\theta_1-16\cos\theta_2-16\cos(\theta_1-\theta_2)$ $C = 48+48e^{i\theta_1}+16e^{i\theta_1}\cos\theta_2+16\cos\theta_2$,

$D = 16+48e^{i\theta_1}+48e^{i\theta_2}+16e^{i(\theta_1+\theta_2)}$, $E = 48+48e^{i\theta_2}+16e^{i\theta_2}\cos\theta_1+16\cos\theta_1$,

$F = 16+48e^{-i\theta_1}+48e^{i\theta_2}+16e^{-i(\theta_1-\theta_2)}$. Using the result (2.15) for d = 2, one can find the resistance between any two sites. For example, the resistance between lattice point 1 and 2 that belong to the same unit cell, we get

$$R_{12}(0,0) = \frac{R}{4\pi^2}\int_{-\pi}^{\pi}d\theta_1\int_{-\pi}^{\pi}d\theta_2\frac{4-2\cos\theta_1-\cos\theta_2-\cos\theta_1\cos\theta_2}{4(3-\cos\theta_1-\cos\theta_2-\cos\theta_1\cos\theta_2)} \quad (3.10)$$

Performing the integrations analytically, we find $R_{12}(0,0) = (1+\pi)R/4\pi$. From the symmetry of the lattice, one can find $R_{12}(0,0) = R_{23}(0,0) = R_{34}(0,0) = R_{14}(0,0) = (1+\pi)R/4\pi$. Moreover, Some other analytical results (in units of $R$) are given below:

$$R_{13}(0,0) = R_{24}(0,0) = \frac{1}{4} + \frac{1}{2\pi},$$

$$R_{11}(1,0) = \frac{1}{8} + \frac{1}{\pi}, \quad R_{14}(1,0) = \frac{3}{8} + \frac{1}{4\pi},$$

$$R_{11}(1,1) = 0.5.$$

*3.2. 2. Union Jack lattice*

The Union Jack lattice is square tiling with each square divided into for triangles from the center point as shown in Fig.3. Each unit cell contains $s = 4$ sites and coordination numbers are 4 and 8. The Ohm's and Kirchhoff's laws applied to the lattice site $\{\mathbf{r}, \alpha\}$ (with $\alpha = 1, 2, 3, 4$) read:

$$\left.\begin{aligned}
RI_1(\mathbf{r}) &= 8V_1(\mathbf{r}) - V_2(\mathbf{r}) - V_2(\mathbf{r}+\mathbf{a}_1) - V_3(\mathbf{r}) - V_3(\mathbf{r}+\mathbf{a}_1) - V_3(\mathbf{r}+\mathbf{a}_2) \\
&\quad - V_3(\mathbf{r}+\mathbf{a}_1+\mathbf{a}_2) - V_4(\mathbf{r}) - V_4(\mathbf{r}+\mathbf{a}_2) \\
RI_2(\mathbf{r}) &= 4V_2(\mathbf{r}) - V_1(\mathbf{r}) - V_1(\mathbf{r}+\mathbf{a}_1) - V_3(\mathbf{r}) - V_3(\mathbf{r}-\mathbf{a}_2) \\
RI_3(\mathbf{r}) &= 8V_3(\mathbf{r}) - V_1(\mathbf{r}) - V_1(\mathbf{r}-\mathbf{a}_1) - V_1(\mathbf{r}-\mathbf{a}_2) - V_1(\mathbf{r}-\mathbf{a}_1-\mathbf{a}_2) \\
&\quad - V_2(\mathbf{r}) - V_2(\mathbf{r}-\mathbf{a}_2) - V_4(\mathbf{r}) - V_4(\mathbf{r}-\mathbf{a}_1) \\
RI_4(\mathbf{r}) &= 4V_4(\mathbf{r}) - V_1(\mathbf{r}) - V_1(\mathbf{r}-\mathbf{a}_2) - V_3(\mathbf{r}) - V_3(\mathbf{r}+\mathbf{a}_1)
\end{aligned}\right\} \quad (3.11)$$

The Fourier transform the Fourier transform of the Laplacian matrix is given by

$$R\mathbf{L}(\theta_1, \theta_2) = \begin{pmatrix} -8 & v_1 & v_1 v_2 & v_2 \\ v_1^* & -4 & v_2 & 0 \\ v_1^* v_2^* & v_2^* & -8 & v_1^* \\ v_2^* & 0 & v_1 & -4 \end{pmatrix}, \text{ where } v_j = 1 + e^{i\theta_j} \ (j=1,2) \quad (3.12)$$

Now calculating the Green's function from equation (2.9) one can use the general expression (2.15) to find the resistance between arbitrary lattice points on the Union Jack resistor network. Some numerical results (in terms of $R$) are given below:

$$R_{12}(0,0) = R_{23}(0,0) = R_{34}(0,0) = R_{14}(0,0) = 0.3615,$$

$$R_{13}(0,0) = 0.2771, \qquad R_{24}(0,0) = 0.5651,$$

$$R_{11}(1,0) = 0.3376, \qquad R_{12}(1,0) = 0.5389,$$

$$R_{22}(1,0) = 0.6094, \qquad R_{11}(1,1) = 0.3891.$$

### 3.3. Lattices in three dimensions

The usual three-dimensional simple, body-centered and face-centered cubic lattices are studied in [5,6], [16] and [17], respectively. In this section, we calculate the resistance on other lattices. In particular on the decorated cubic and base-centered lattices.

*3.3.1. The decorated cubic lattice*

The so-called decorated cubic lattice is a simple cubic lattice with one site inserted on each edge as shown in Fig. 4. The unit cell contains $s = 4$ sites labeled by $\alpha = 1,2,3,4$ and the degree or coordination number of all sites is 2 or 6. Using Ohm's and Kirchhoff's law's the equation for the currents for each lattice site in one unit cell can be written as

$$\left.\begin{aligned}
I_1(\mathbf{r})R &= 6V_1(\mathbf{r}) - V_2(\mathbf{r}) - V_2(\mathbf{r}-\mathbf{a}_1) - V_3(\mathbf{r}) - V_3(\mathbf{r}-\mathbf{a}_2) - V_4(\mathbf{r}) - V_4(\mathbf{r}-\mathbf{a}_3) \\
I_2(\mathbf{r})R &= 2V_2(\mathbf{r}) - V_1(\mathbf{r}) - V_1(\mathbf{r}+\mathbf{a}_1) \\
I_3(\mathbf{r})R &= 2V_3(\mathbf{r}) - V_1(\mathbf{r}) - V_1(\mathbf{r}+\mathbf{a}_2) \\
I_4(\mathbf{r})R &= 2V_4(\mathbf{r}) - V_1(\mathbf{r}) - V_1(\mathbf{r}+\mathbf{a}_3)
\end{aligned}\right\} \quad (3.13)$$

After changing the variables $\mathbf{k}\mathbf{a}_i = \theta_i \ (i = 1,2,3)$, the Fourier transform of the Laplacian matrix is given by

$$R\mathbf{L}(\theta_1,\theta_2,\theta_3) = \begin{bmatrix} -6 & v_1 & v_2 & v_3 \\ v_1^* & -2 & 0 & 0 \\ v_2^* & 0 & -2 & 0 \\ v_3^* & 0 & 0 & -2 \end{bmatrix} \quad \text{where } v_j = 1 + e^{-i\theta_j}, \ j = 1,2,3. \quad (3.14)$$

and the corresponding lattice Green's function :

$$\mathbf{G}(\theta_1,\theta_2,\theta_3) = \frac{1}{Det(\mathbf{L}(\theta_1,\theta_2,\theta_3))} \begin{pmatrix} 8 & 4v_1 & 4v_2 & 4v_3 \\ 4v_1^* & A & 2v_1^*v_2 & 2v_1^*v_3 \\ 4v_2^* & 2v_1v_2^* & B & 2v_2^*v_3 \\ 4v_3^* & 2v_1v_3^* & 2v_2v_3^* & C \end{pmatrix} \quad (3.15)$$

where $Det(\mathbf{L}(\theta_1,\theta_2,\theta_3))R = 8(3 - \cos\theta_1 - \cos\theta_2 - \cos\theta_3)$, $A = 16 - \cos\theta_2 - \cos\theta_3$, $B = 16 - \cos\theta_1 - \cos\theta_3$, $C = 16 - \cos\theta_1 - \cos\theta_2$.

One can note that the determinant of the Laplacian $\mathbf{L}(\theta_1,\theta_2)$ is eight times the denominator of the integrand in the resistance formula for the simple cubic lattice [5,6], so that the resistance on the decorated simple cubic lattice $R_{\alpha\beta}(\ell_1,\ell_2,\ell_3)$ can be expressed in terms of the resistances $R_{sc}(\ell_1,\ell_2,\ell_3)$ on the simple cubic lattice. The resistance between the origin and the site $(\ell_1,\ell_2,\ell_3)$ on the simple cubic lattice is given by

$$R_{sc}(\ell_1,\ell_2,\ell_3) = \frac{R}{(2\pi)^3} \int_{-\pi}^{\pi} d\theta_1 \int_{-\pi}^{\pi} d\theta_2 \int_{-\pi}^{\pi} d\theta_3 \frac{1 - \cos(\ell_1\theta_1 + \ell_2\theta_2 + \ell_3\theta_3)}{3 - \cos\theta_1 - \cos\theta_2 - \cos\theta_3} \quad (3.16)$$

The results for the resistance $R_{\alpha\beta}(\ell_1,\ell_2,\ell_3)$ on a decorated simple cubic lattice in terms of the resistances $R_{sc}(\ell_1,\ell_2,\ell_3))$ on an usual simple cubic lattice are given in Appendix. Some results for resistances are presented below:

$$R_{12}(0,0,0) = R_{13}(0,0,0) = R_{14}(0,0,0) = \frac{2}{3},$$

$$R_{24}(0,0,0) = 1.1975, \quad R_{12}(1,0,0) = 1.0864,$$

$$R_{11}(1,1,0) = 0.7902, \quad R_{11}(1,1,1) = 0.8367,$$

$$R_{22}(1,0,0) = 1.2098$$

*3.3.2. Base-centered cubic lattice*

The base-centered cubic lattice is a three dimensional lattice. Besides the resistors on the edges of the cube there are resistors between the centers of its two bases to the corners of these bases as shown in Fig.5 . Each unit cell consists of two lattice points. Applying the Ohm's and Kirchhoff's law's one can write the current at each lattice point in one unit cell as

$$\left.\begin{array}{l} RI_1(\mathbf{r}) = \sum_{i=1}^{3}\left(2V_1(\mathbf{r})-V_1(\mathbf{r}+\mathbf{a}_i)-V_1(\mathbf{r}-\mathbf{a}_i)\right) + 4V_1(\mathbf{r}) - V_2(\mathbf{r}) - V_2(\mathbf{r}-\mathbf{a}_1) \\ \qquad - V_2(\mathbf{r}-\mathbf{a}_2) - V_2(\mathbf{r}-\mathbf{a}_1-\mathbf{a}_2) \\ RI_2(\mathbf{r}) = 4V_2(\mathbf{r}) - V_1(\mathbf{r}) - V_1(\mathbf{r}+\mathbf{a}_1) - V_1(\mathbf{r}+\mathbf{a}_2) - V_1(\mathbf{r}+\mathbf{a}_1+\mathbf{a}_2) \end{array}\right\} \quad (3.17)$$

Therefore, the Fourier transform of the Laplacian matrix is given by

$$R\mathbf{L}(\theta_1,\theta_2,\theta_3) = \begin{pmatrix} -10 + 2\cos\theta_1 + 2\cos\theta_2 + 2\cos\theta_3 & (1+e^{-i\theta_1})(1+e^{-i\theta_2}) \\ (1+e^{i\theta_1})(1+e^{i\theta_2}) & -4 \end{pmatrix} \quad (3.18)$$

and the corresponding Green's function :

$$\mathbf{G}(\theta_1,\theta_2,\theta_3) = \frac{1}{Det(\mathbf{L})} \begin{pmatrix} 4 & (1+e^{-i\theta_1})(1+e^{-i\theta_2}) \\ (1+e^{i\theta_1})(1+e^{i\theta_2}) & 10 - 2\cos\theta_1 - 2\cos\theta_2 - 2\cos\theta_3 \end{pmatrix} \quad (3.19)$$

where $Det(\mathbf{L})R = 36 - 12\cos\theta_1 - 12\cos\theta_2 - 4\cos\theta_1\cos\theta_2 - 8\cos\theta_3$.

Now, the resistance between any two site can be calculated from Eq.(2.15). Below we present some numerical results in units of $R$

$R_{12}(0,0,0) = 0.334, \quad R_{11}(1,0,0) = 0.2128,$

$R_{12}(1,0,0) = 0.4167, \quad R_{11}(1,1,0) = 0.2465,$

$R_{11}(1,1,1) = 0.2739, \quad R_{22}(1,0,0) = 0.5414$

## 4. Conclusions

In this work, we have calculated the resistance between any two nodes on the infinite ladder lattice, the modified square , Union Jack, decorated simple cubic and base-centered cubic lattices using the general Green's function method[16]. For the decorated simple cubic lattice we have expressed the resistances between two arbitrary sites in terms of the resistances on a simple cubic lattice of resistors. We have presented some analytical and numerical results near the lattices origins.

**Appendix. Relation between the decorated simple cubic lattice and the usual simple cubic lattice of resistor networks**

In this appendix we summarize the results for the resistance $R_{\alpha\beta}(\ell_1,\ell_2,\ell_3)$ on a decorated simple cubic lattice in terms of the resistances $R_{sc}(\ell_1,\ell_2,\ell_3)$) on an usual simple cubic lattice:

$$R_{11}(\ell_1,\ell_2,\ell_3) = 2R_{sc}(\ell_1,\ell_2,\ell_3) \qquad \text{A.1}$$

$$R_{22}(\ell_1,\ell_2,\ell_3) = \frac{2R}{3} + 4R_{sc}(\ell_1,\ell_2,\ell_3) - \frac{1}{2}\left\{\begin{array}{l} R_{sc}(\ell_1,\ell_2+1,\ell_3) + R_{sc}(\ell_1,\ell_2-1,\ell_3) \\ +R_{sc}(\ell_1,\ell_2,\ell_3+1) + R_{sc}(\ell_1,\ell_2,\ell_3-1) \end{array}\right\} \qquad \text{A.2}$$

$$R_{33}(\ell_1,\ell_2,\ell_3) = \frac{2R}{3} + 4R_{sc}(\ell_1,\ell_2,\ell_3) - \frac{1}{2}\left\{\begin{array}{l} R_{sc}(\ell_1+1,\ell_2,\ell_3) + R_{sc}(\ell_1-1,\ell_2,\ell_3) \\ +R_{sc}(\ell_1,\ell_2+1,\ell_3) + R_{sc}(\ell_1,\ell_2-1,\ell_3) \end{array}\right\} \qquad \text{A.3}$$

$$R_{44}(\ell_1,\ell_2,\ell_3) = \frac{2R}{3} + 4R_{sc}(\ell_1,\ell_2,\ell_3) - \frac{1}{2}\left\{\begin{array}{l} R_{sc}(\ell_1+1,\ell_2,\ell_3) + R_{sc}(\ell_1-1,\ell_2,\ell_3) \\ +R_{sc}(\ell_1,\ell_2,\ell_3+1) + R_{sc}(\ell_1,\ell_2,\ell_3-1) \end{array}\right\} \qquad \text{A.4}$$

$$R_{12}(\ell_1,\ell_2,\ell_3) = \frac{R}{3} + R_{sc}(\ell_1,\ell_2,\ell_3) + R_{sc}(\ell_1+1,\ell_2,\ell_3) \qquad \text{A.5}$$

$$R_{13}(\ell_1,\ell_2,\ell_3) = \frac{R}{3} + R_{sc}(\ell_1,\ell_2,\ell_3) + R_{sc}(\ell_1,\ell_2+1,\ell_3) \qquad \text{A.6}$$

$$R_{14}(\ell_1,\ell_2,\ell_3) = \frac{R}{3} + R_{sc}(\ell_1,\ell_2,\ell_3) + R_{sc}(\ell_1,\ell_2,\ell_3+1) \qquad \text{A.7}$$

$$R_{23}(\ell_1,\ell_2,\ell_3) = \frac{2R}{3} + \frac{1}{2}\left\{\begin{array}{l} R_{sc}(\ell_1,\ell_2,\ell_3) + R_{sc}(\ell_1-1,\ell_2,\ell_3) \\ +R_{sc}(\ell_1,\ell_2+1,\ell_3) + R_{sc}(\ell_1-1,\ell_2+1,\ell_3) \end{array}\right\} \qquad \text{A.8}$$

$$R_{24}(\ell_1,\ell_2,\ell_3) = \frac{2R}{3} + \frac{1}{2}\left\{\begin{array}{l} R_{sc}(\ell_1,\ell_2,\ell_3) + R_{sc}(\ell_1-1,\ell_2,\ell_3) \\ +R_{sc}(\ell_1,\ell_2,\ell_3+1) + R_{sc}(\ell_1-1,\ell_2,\ell_3+1) \end{array}\right\} \qquad \text{A.9}$$

$$R_{34}(\ell_1,\ell_2,\ell_3) = \frac{2R}{3} + \frac{1}{2}\left\{\begin{array}{l} R_{sc}(\ell_1,\ell_2,\ell_3) + R_{sc}(\ell_1,\ell_2+1,\ell_3) \\ +R_{sc}(\ell_1,\ell_2,\ell_3-1) + R_{sc}(\ell_1,\ell_2+1,\ell_3-1) \end{array}\right\} \qquad \text{A.10}$$

# Figure captions

**Fig.1.** Infinite strip of the triangular lattice of resistors as an example of a one dimensional network

**Fig.2.** The modified square lattice of the resistor network. Sites within a unit cell are labeled as shown.

**Fig.3.** The Union Jack( tetrakis) lattice of the resistor network. Sites within a unit cell are labeled as shown.

**Fig.4.** The so-called decorated simple cubic lattice. Sites within a unit cell are labeled as shown.

**Fig. 5.** Base-centered lattice structure of the resistor network. Each unit cell consists of two sites.

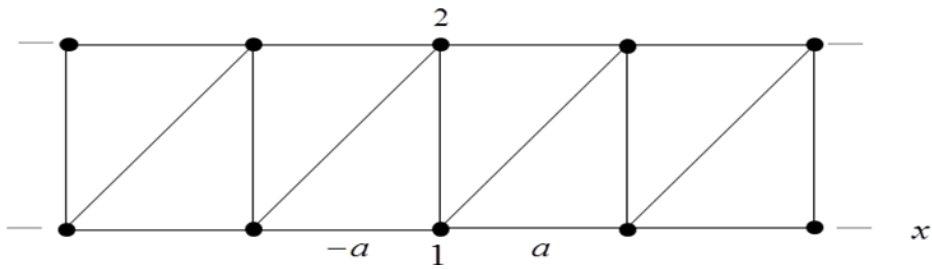

**Fig.1**

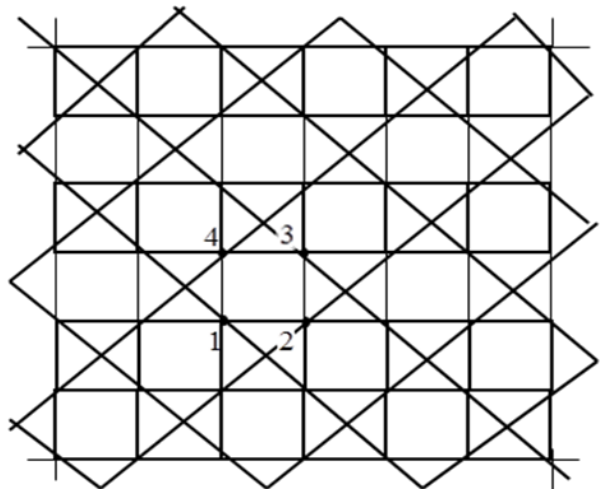

**Fig.2**

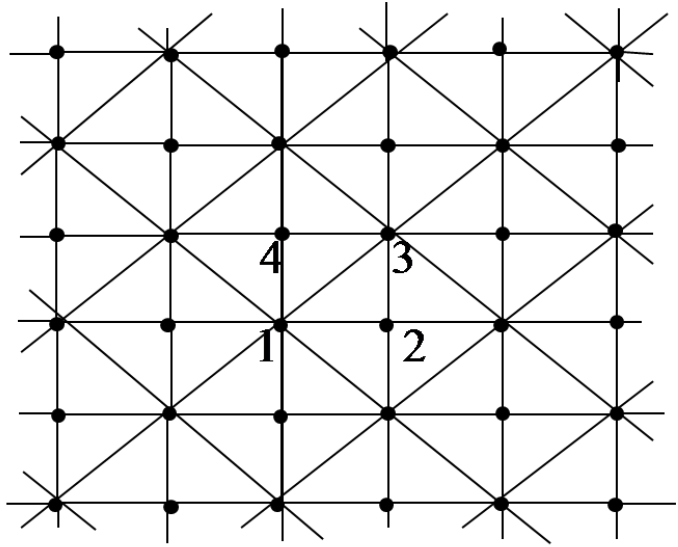

**Fig.3**

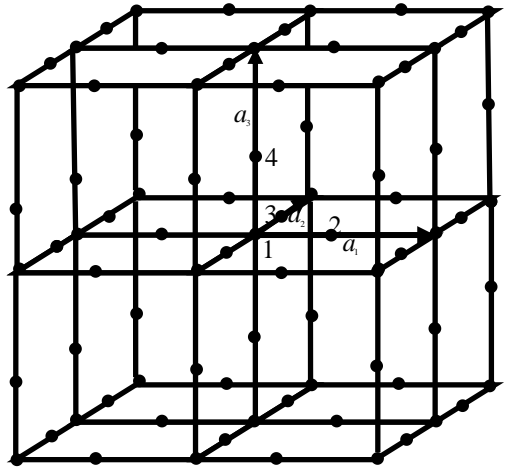

**Fig.4**

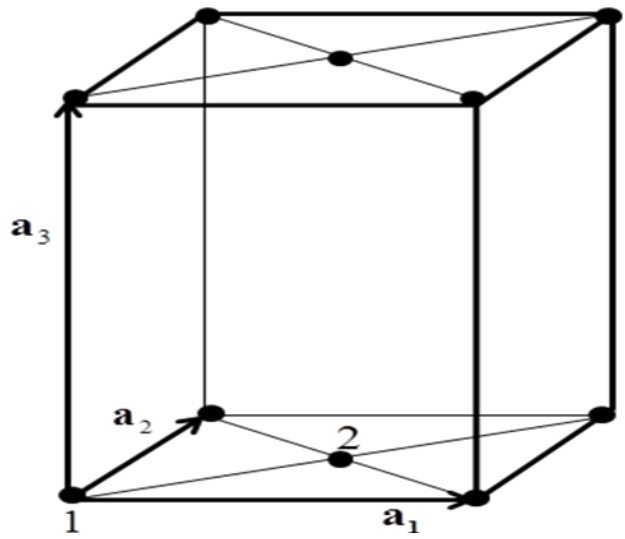

**Fig. 5**